\documentclass[12pt]{article}
\usepackage{amsmath,eucal}
\newcommand{\R}{\mbox{\bf R}}
\newcommand{\C}{\mbox{\bf C}}
\begin{document}
\title{Generalized Weierstrass representation for surfaces and Lax-Phillips
scattering theory for automorphic functions}
\author{Vadim V. Varlamov\\
{\small\it Computer Division, Siberian State Industrial University,}\\
{\small\it Novokuznetsk 654007, Russia}}
\date{}
\maketitle
\begin{abstract}
Relation between generalized Weierstrass representation for conformal
immersion of generic surfaces into three-dimensional space and Lax-Phillips
scattering theory for automorphic functions is considered.
\end{abstract}
\vspace{0.5cm}
It is well-known that Poincare plane $\Pi$, i.e., the upperhalf plane
$$y>0,\quad-\infty<x<\infty,\quad z=x+iy$$
be the model of Lobachevsky geometry, where the role of motion group played
the group $G=SL(2,\R)$ of fractional linear transformations
\begin{equation}\label{e1}
z\longrightarrow zg=\frac{az+b}{cz+d},\quad g=\begin{pmatrix} a & b\\
c & d\end{pmatrix},
\end{equation}
where $a,b,c,d\in\R,\;ad-bc=1$.

The group $SL(2,\R)$ has a great number of so-called {\it discrete subgroups}.
The sugroup $\Gamma$ is called discrete if the identical transformation
is isolate from the other transformations $\gamma\in\Gamma$. For example,
{\it a modular group} consisting of transformations with integer
$a,b,c,d$ is discrete subgroup. Further, {\it a fundamental domain} $F$ of
discrete subgroup $\Gamma$ be an any domain  on Poincare plane such that
the every point of $\Pi$ may be transfered into a closing $\bar{F}$ of
domain $F$ by means of some transformation $\gamma\in\Gamma$, at the same
time no there exists the point from $F$ which transfered to the other point
of $F$ by such transformation. The function $f$ defined on $\Pi$ is called
{\it automorphic} with reference to discrete subgroup $\Gamma$ if
$$f(\gamma z)=f(z),\quad\gamma\in\Gamma.$$

Further, generalized Weierstrass representation for surfaces was proposed
by Konopelchenko in 1993 \cite{1,2} is defined by the following formulae
\begin{eqnarray}
X^1&+&iX^2=i\int_\epsilon(\bar{\psi}^2dz^\prime-\bar{\varphi}^2d\bar{z}^\prime),
\nonumber\\
X^1&-&iX^2=i\int_\epsilon(\varphi^2dz^\prime-\psi^2d\bar{z}^\prime),\nonumber\\
X^3&=&-\int_\epsilon(\psi\bar{\varphi}dz^\prime+\varphi\bar{\psi}d\bar{z}^
\prime),
\label{e2}
\end{eqnarray}
where $\epsilon$ is arbitrary curve in $\C$, $\psi$ and $\varphi$ are 
complex-valued functions on variables $z,\bar{z}\in\C$ satisfying to the
linear system (two-dimensional Dirac equation):
\begin{eqnarray}
\psi_z&=&U\varphi,\nonumber\\
\varphi_{\bar{z}}&=&-U\psi,\label{e3}
\end{eqnarray}
where $U(z,\bar{z})$ is a real-valued function. If to interpret the
functions $X^i(z,\bar{z})$ as coordinates in a space $\R^{3,0}$, then the
formulae (\ref{e2}), (\ref{e3}) define a conformal immersion of surface
into $\R^{3,0}$ with induced metric 
$$ds^2=D(z,\bar{z})^2dzd\bar{z},\quad D(z,\bar{z})=\left|\psi(z,\bar{z})
\right|^2+\left|\varphi(z,\bar{z})\right|^2,
$$
at this the Gaussian and mean curvature are
\begin{equation}\label{e4}
K=-\frac{4}{D^2}[\log D]_{z\bar{z}},\quad H=\frac{2U}{D}.
\end{equation}

Let us consider a closed surface with genus $>1$, and let $F:\Sigma
\longrightarrow\R^{3,0}$ be an immersion of surface with genus $>1$ given by
(\ref{e2})-(\ref{e3}). It is well-known that every closed oriented surface
$\Sigma$ with positive genus is uniformizable:
$$\rho: M\longrightarrow\Sigma,$$
where a surface $M$ is conformal covering. Hence it immediately follows
that a factor-space $M/\Gamma$ is conformally equivalent to the surface
$\Sigma$, where $\Gamma$ is a discrete subgroup of a group of isometries
of $M$. In our case a space $M$ is isometric to the Poincare plane $\Pi$.
The group of isometries of $\Pi$ is the group $G=SL(2,\R)$, the 
transformations of which are defined by (\ref{e1}).

According to \cite{3} (Proposition 4) we have that a surface $\Sigma$ with
genus $>1$ immersing into $\R^{3,0}$ by formulas (\ref{e2})-(\ref{e3})
is conformally equivalent to a surface $\Pi/\Gamma$, where $\Gamma$ is a
discrete subgroup of $SL(2,\R)$. The functions $\psi$ and $\varphi$, the
metric tensor $D(z)^2$ and potential $U(z)$, are transformed by elements
of $\Gamma$ as follows
\begin{eqnarray}
\psi(\gamma(z))&=&(c\bar{z}+d)\psi(z),\nonumber\\
\varphi(\gamma(z))&=&(cz+d)\varphi(z),\nonumber\\
D(\gamma(z))&=&|cz+d|^2D(z),\nonumber\\
U(\gamma(z))&=&|cz+d|^2U(z).\nonumber
\end{eqnarray}
Hence it immediately follows from (\ref{e4}) that
\begin{equation}\label{e5}
H(\gamma(z))=H(z),\quad\gamma\in\Gamma.
\end{equation}
Therefore, {\it the mean curvature is automorphic function}.

Further, follows to \cite{4} let us consider the discrete subgroup
$\Gamma\subset SL(2,\R)$ satisfying to the following requirements:
\begin{description}
\item[1.] A space $SL(2,\R)/\Gamma$ is noncompact.
\item[2.] $\Gamma$  contains the only one parabolic subgroup.
\end{description}
The fundamental domain $F_\Gamma=F$ for the group $\Gamma$ choosing as
follows
\begin{description}
\item[a.] $F$ lies on a strip $-X<x<X,\;y>Y>0$.
\item[b.] Intersection $F\cap\{y>d\}$ at the some $d,\;d>1$, is 
coincide with a strip $-X_1<x<X_1,\;y>d$.
\item[c.] A boundary of $F$ is smooth and consists of geodesic
segments with a finite number of corner points.
\end{description}

In Hilbert space $L_2(F,d\mu)$, where $d\mu=y^{-2}dxdy$ is a measure,
consider a symmetric operator $L$ defined by the differential expression
\begin{equation}\label{e6}
L=-y^2\left(\frac{\partial^2}{\partial x^2}+\frac{\partial^2}{\partial y^2}
\right)-\frac{1}{4}
\end{equation}
on all sufficiently smooth and uniformly restricted in $F$ the functions
$H$, which satisfying to automorphity condition (\ref{e5}). A spectrum of
the operator $L$ consists of finite set of own values: $\lambda_0=
-\frac{\displaystyle 1}{\displaystyle 4}$ (this own number corresponds to
the unit representation of group $SL(2,\R)$), the numbers $\lambda_l=-
\mu^2_l,\;l=1,2,\ldots, N$ are belong to ($-\frac{\displaystyle 1}{
\displaystyle 4},0$) (additional series), the set of positive own values
$\lambda_l,\;l=N+1, N+2,\ldots,\;\lambda_l\in(0,\infty)$ (basic series),
and the branch of absolutely continuous spectrum $\lambda=k^2$ on
$[0,\infty]$.
\begin{sloppypar}
If we assume that functions $H$ compose the basis of Hilbert space
$L_2(F,d\mu)$, then {\it the automorphic wave equation} may be written as
\end{sloppypar}
\begin{equation}\label{e7}
H_{tt}+LH=0,
\end{equation}
where operator $L$ has the form (\ref{e6}). This equation naturally defines
a group $\mathcal{V}_t$ of transformations (smooth and finite) of Cauchy
data $\mathcal{U}(z,t)=\begin{pmatrix} u(z,t)\\ \frac{\partial}{\partial t}
u(z,t)\end{pmatrix}$, the action of which expressed by the formula
$$\mathcal{U}(z,t)=\mathcal{V}_t\mathcal{U}(z,0).$$
The group $\mathcal{V}_t$ has orthogonal in- and out-spaces $\mathcal{D}_-,\;
\mathcal{D}_+$ are satisfying to conditions
\begin{eqnarray}
&&1)_-\quad\mathcal{V}_t\mathcal{D}_-\subset\mathcal{D}_-,\;t<0,\nonumber\\
&&1)_+\quad\mathcal{V}_t\mathcal{D}_+\subset\mathcal{D}_+,\;t>0,\nonumber\\
&&2)\phantom{{}_+}\quad\underset{t<0}{\bigcap}\mathcal{V}_t\mathcal{D}_-=
\underset{t>0}{\bigcap}\mathcal{V}_t\mathcal{D}_+=0,\nonumber\\
&&3)\phantom{{}_+}\quad\underset{t>0}{\bigcup}\mathcal{V}_t\mathcal{D}_-=
\underset{t>0}{\bigcup}\mathcal{V}_t\mathcal{D}_+,\nonumber\\
&&4)\phantom{{}_+}\quad\mathcal{D}_-\perp\mathcal{D}_+.\nonumber
\end{eqnarray}
These conditions allow to apply the Lax-Phillips framework \cite{5} and to
find generalized own functions $e(z)$ of automorphic wave equation
(\ref{e7}) which are expressed via the Eisenstein series, and also to define
the spectrum representation of operator $L$ and scattering matrix.  

\end{document}